\documentclass[a4paper]{article}

\usepackage{icrc2013}
\usepackage[english]{babel} 
\usepackage{amsmath,amssymb,amscd}

\title{High-energy fluxes of atmospheric neutrinos}

\shorttitle{Fluxes of atmospheric neutrinos}

\authors{
T.~S.~Sinegovskaya$^{1}$,
E.~V.~Ogorodnikova$^{2}$,
S.~I.~Sinegovsky$^{2}$
}

\afiliations{
$^1$ Irkutsk State Railway University, 664074 Irkutsk, Russia \\ 
$^2$ Irkutsk State University, 664003 Irkutsk, Russia  \\ 
}

\email{sinegovsky@api.isu.ru}

\abstract{High-energy  neutrinos from decays of mesons, produced in collisions of cosmic
ray  particles  with  air  nuclei,  form  unavoidable  background for detection of astrophysical  neutrinos.  More  precise  calculations of the high-energy neutrino spectrum  are  required  since  measurements  in  the IceCube experiment reach the intriguing  energy  region  where  a  contribution  of the prompt neutrinos and/or astrophysical  ones  should  be discovered. Basing on the referent hadronic models
QGSJET II-03, SIBYLL  2.1, we calculate high-energy spectra, both of the muon   and  electron  atmospheric  neutrinos,  averaged  over  zenith-angles.  The computation  is  made  using  three  parameterizations of cosmic ray spectra which include  the  knee  region.  All  calculations  are  compared with the atmospheric
neutrino  measurements  by  Frejus  and IceCube.  The prompt  neutrino flux predictions obtained with the  quark-gluon  string model (QGSM) for the charm production by Kaidalov \& Piskunova 
do not contradict  to  the  IceCube  measurements and upper limit on the astrophysical muon neutrino flux.
Neutrino flavor ratio, $\phi_{\nu_\mu}/\phi_{\nu_e}$, extracted from IceCube data  decreases in the energy range $0.1 - 5$ TeV  energy  contrary to that one might expect from the conventional neutrino flux. Presumable reasons of such behavior are: i) early arising contribution from decays of charmed particle, differing from predictions of present models, ii)  revealed  diffuse  flux  of astrophysical electron neutrinos. 
The  likely diffuse flux of astrophysical neutrinos related to the PeV neutrino events, detected in the IceCube experiment, leads to a decrease  of  the flavor ratio at the energy below $10$ TeV, that is in qualitative  agreement  with  a rough  approximation for the flavor ratio obtained from the IceCube data. 
}


\begin{document}
\maketitle

\section{Introduction}

High-energy neutrinos produced in decays of muons, pions, kaons, and charmed particles
of the extensive air shower induced by cosmic rays in the Earth atmosphere,  form an unavoidable 
background for the detection of astrophysical neutrinos. Search of extraterrestrial neutrino sources is a challenge to resolve which large-scale neutrino telescopes, NT200+~\cite{nt200-08}, 
IceCube~\cite{icecube11, icecube40_lim, IC_dif2011},  ANTARES~\cite{antares11} are designed. 
The high-energy atmospheric neutrinos became accessible  to the  experimental studies only last years.
By now, the energy spectrum of high-energy atmospheric muon neutrinos has been measured  in  three experiments: Frejus~\cite{frejus} at energies up to 1 TeV, AMANDA-II~\cite{amanda10} in the energy range 
 $1-100$ TeV, and IceCube40~\cite{icecube11} in the range $100$ GeV -- $400$ TeV. Recently the IceCube  presents results for the electron neutrino spectrum  measured in the energy range $\sim 80$ GeV -- $6$ TeV~\cite{ice79_nue}. Thus one has a possibility to obtain the neutrino flavour ratio from IceCube experiment and to compare this one with  predictions. 

The increasing with the energy contribution of charmed particle decays to the neutrino flux becomes the  source of the large uncertainty at energies above $100$ TeV. Thus the comparison of the calculation for various hadron-interaction models with neutrino spectrum measurements is of interest, despite large statistical and systematic experimental errors in  the high-energy region. Here we calculate atmospheric neutrino fluxes at energies $10^2-10^7$ GeV for zenith angles from $0^\circ$ to $90^\circ$ as well as the angle averaged  spectrum with the use of high-energy hadronic interaction models QGSJET II-03~\cite{qgsjet2} and SIBYLL 2.1~\cite{sibyll}, which are widely employed to simulate extensive air showers with the Monte Carlo method, and were also used to compute the cosmic-ray hadron and muon fluxes~\cite{kss08, kss13}. 

The calculation has been performed for three parameterizations of the experimentally measured spectrum and the composition of primary cosmic rays (PCR) in the  energy range comprising the knee: 1)  the model by Zatsepin \& Sokolskaya~\cite{ZS3C}, 2) the modified multi-knee model  by Bindig, Bleve and Kampert~\cite{BK2011}, and 3) the novel model of primary spectrum  by Gaisser~\cite{HiG2012}, based on assumption that there are three classes  of CR sources: i) Galactic supernova remnants, ii) Galactic high-energy sources diffent from the former, iii) extragalactic astrophysical objects.

\section{Fluxes of atmospheric muon neutrinos \label{sec:numu}} 

The calculation is performed on the basis of the method~\cite{ns2000} of solution of the hadronic cascade equations in the atmosphere,  which takes into account nonscaling behavior of inclusive particle production cross-sections, the rise of total inelastic hadron-nuclei cross-sections, and the nonpower law primary spectrum (see also~\cite{kss08}). Along with major sources of the muon neutrinos, $\pi_{\mu2}$ and  $K_{\mu2}$ decays, we consider three-particle semileptonic decays, $K^{\pm}_{\mu3}$, $K^{0}_{\mu3}$,  the contribution originated from decay chains   $K\rightarrow\pi\rightarrow\nu_\mu$ ($K^0_S\rightarrow \pi^+\pi^-$, $K^\pm \rightarrow \pi^\pm \pi ^0$), as well as small fraction from the muon decays. 
The sources of the conventional $\nu_e$'s are three-particle decays of kaons $K^{\pm}_{e3}$, $K^{0}_{e3}$ and also $\mu_{e3}$ decay.

\begin{figure}[!t]
 \centering  \vskip -0.0 cm
\includegraphics[width=80 mm]{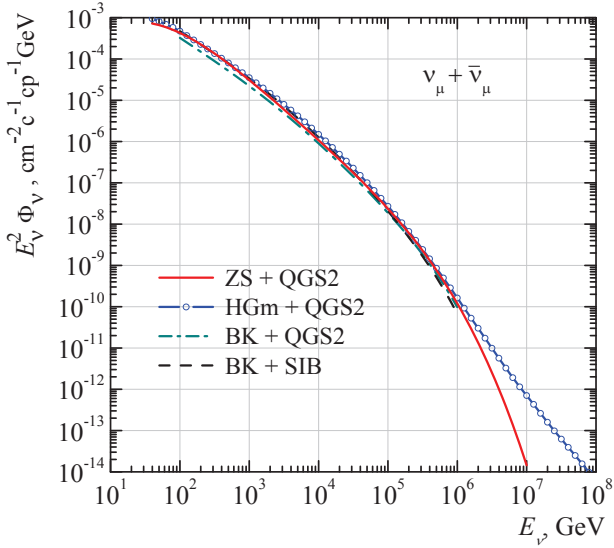} 
\vskip -0.2 cm   
\caption{Comparison $\nu_{\mu}+\bar\nu_{\mu}$ fluxes calculated for 3 PCR models.}
\label{compar_3pcr}  
\end{figure}
\begin{figure}[!t]
  \centering  
\includegraphics[ width=75 mm]{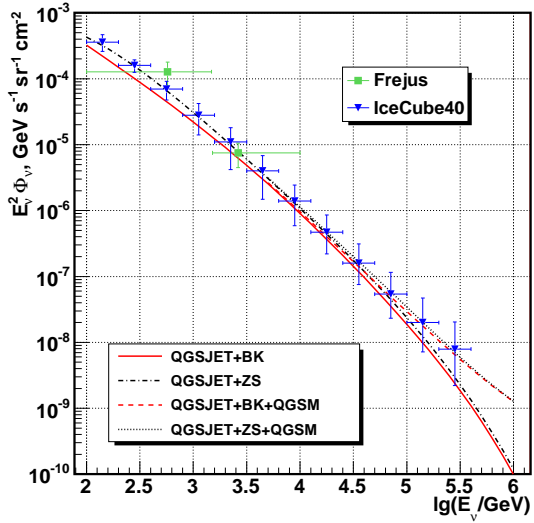} 
 \caption{Fluxes of the conventional and prompt muon neutrinos ($\nu_\mu + \bar{\nu}_\mu$) calculated with  QGSJET II-03 hadronic model for ZS and BK primary spectrum.} 
\label{numu-qgs2} 
 \end{figure} 
\begin{figure}[!t]  
  \centering \vskip -0.2 cm
\includegraphics[width=75 mm]{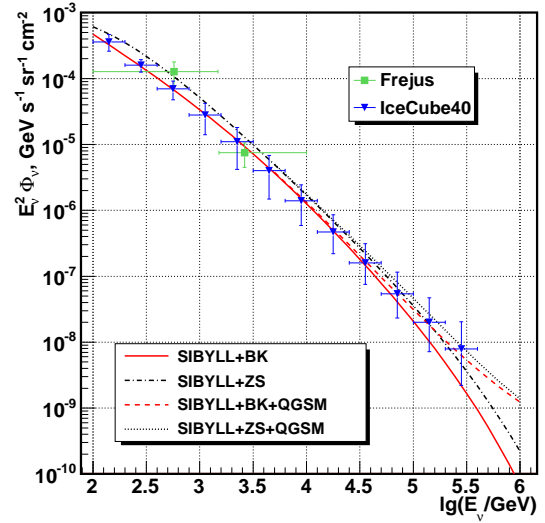} 
 \caption{Fluxes of the conventional and prompt (QGSM) muon neutrinos ($\nu_\mu + \bar{\nu}_\mu$) calculated with the  SIBYLL 2.1.} 
  \label{numu_sib} 
 \end{figure}
As the primary cosmic ray spectra and composition in wide energy range folowing models are used: 1)  the model by Zatsepin \& Sokolskaya (ZS), 2) the modified multi-knee model  by Bindig, Bleve and Kampert (BK) based on KASCADE data~\cite{KASCADE05} and the  polygonato model by H$\ddot{\text o}$randel ~\cite{polygon}, and 3) the novel CR  approximation by Gaisser~\cite{HiG2012}, a version for the mixed CR population 3 (HGm here). 

The model by Zatsepin and Sokolskaya~\cite{ZS3C}  describes well data of the ATIC2 direct measurements~\cite{atic2_07, atic2_09} in the range  $10–10^5$  GeV and gives a motivated extrapolation of these data up to  $100$ PeV  -- the energy region, for which the cosmic ray spectrum and composition is reconstructed based on the measured characteristics of EAS. 
The ZS proton spectrum at $E\gtrsim 10^6$ GeV is compatible with KASCADE data~\cite{KASCADE05} as well the helium  one within the range of the KASCADE spectrum obtained with the usage of hadronic models QGSJET01 and SIBYLL, and well agree with the HGm up to $1$ PeV.

Comparison of the the muon neutrino fluxes calculated with three recent primary spectrum model models  (Fig.~\ref{compar_3pcr}) shows that they are rather close each other up to $1$ PeV. 
Comparison of calcultion results with IceCube experimental data is shown in 
Figs.~\ref{numu-qgs2}, \ref{numu_sib}.  The difference of muon neutrino flux predictions resulted from the primary cosmic ray spectra becomes apparent at high neutrino energies: the flux obtained with QGSJET II for  ZS spectrum at $2$ PeV is less by a third  of the flux for HGm spectrum.  More results of the muon meutrino calculations, including a comparison with the AMANDA measurments, were presented in Ref.~\cite{SPS11,PSS12}. 
 
The calculation of conventional $\nu_\mu+\bar\nu_\mu$  fluxes averaged over zenith angles
as compared with Frejus~\cite{frejus} and IceCube~\cite{icecube11} measurement data is shown in Figs.~\ref{numu-qgs2},  \ref{numu_sib}. 
 In Fig.~\ref{numu-qgs2} curves displays  the conventional $\nu_\mu+\bar\nu_\mu$  energy spectrum   calculated with usage of QGSJET{\text -}II model  for BK primary spectra and composition (solid line) as well as for ZS one (dashed). 
\begin{table}[!b]
   \begin{center}  \vspace{0.2cm} 
   \begin{tabular}{|l|c|}  \hline 
     Model & $E_\nu^2\phi_\nu$, GeV\,(cm$^{2}$\,s\,sr)$^{-1}$  \\ \hline
  conventional $\nu_\mu+\bar\nu_\mu$: &  $200-400$ TeV  \\ 
  ZS+SIBYLL 2.1   &  $9.4\times 10^{-9} - 2.2\times 10^{-9}$ \\ [3pt]
  ZS+QGSJET-II-03 &  $6.1\times 10^{-9} - 1.3\times 10^{-9}$ \\ [3pt]  
  HGm+QGSJET-II-03 &  $6.7\times 10^{-9} - 1.5\times 10^{-9}$   \\ \hline  
   prompt $\nu_\mu+\bar\nu_\mu$:  & $200-400$ TeV \\ 
  RQPM \cite{bnsz89}    & $2.1 \cdot 10^{-8} - 9.3 \cdot 10^{-9}$ \\[3pt] 
  QGSM \cite{bnsz89}  & $5.1 \times 10^{-9} - 2.2 \cdot 10^{-9}$  \\ \hline 
  IC59 limit~\cite{ice59_lim}  & $1.4 \cdot 10^{-8}$   \\
     ($35$ TeV -– $7$ PeV)  & \\
  ANTARES limit~\cite{antares11}   & $5.3 \times 10^{-8}$   \\  
     ($20$ TeV -– $2.5$ PeV) &                                \\ \hline  
   \end{tabular}  
\caption{ Atmospheric neutrino flux and upper limit for diffuse ($\nu_\mu+\bar\nu_\mu$) flux obtained with neutrino telescopes.} 
\label{tab_1}
   \end{center}
  \end{table}  
The prompt neutrino flux was calculated~\cite{bnsz89} using the quark-gluon string model (QGSM)   by Kaidalov \& Piskunova~\cite{KP} to describe the charmed particle production in nucleon-nucleus collisons at high enrgies. This clculation was peformed with NSU primary spectrum~\cite{NSU}, therefore they can serve here as upper limits for the prompt neutrino flux due to RQPM  or  QGSM.  
Notice that estimate of the prompt neutrino flux obtained with the dipole model~\cite{DM}  is in close agreement to the QGSM prediction~\cite{bnsz89}  above  $1$ PeV. 
 The prompt neutrino flux due to QGSM in the energy range $5$ TeV $\leq E_{\nu} \leq 5\cdot 10^3$ TeV was approximated by the expression  
\begin{equation}\label{pms}
\Phi_{\nu}^{\rm pr}(E_{\nu})
 = A(E_\nu/E_1)^{-3.01}[1+(E_\nu/E_1)^{-2.01}]^{-0.165},
\end{equation}
 where $A=1.19\cdot 10^{-18}\,\mathrm{(GeV\,cm^{2}\,s\,sr)^{-1}}$, $E_1=100$ TeV.
 
Calculated atmospheric muon neutrino fluxes for the energy range $200-400$ TeV are presented in Table~\ref{tab_1} along with upper limits on the  astrophysical muon neutrino diffuse flux obtained with the IceCube59 ~\cite{ice59_lim} and ANTARES~\cite{antares11}. Note that calculated total atmospheric neutrino flux -- sum of the prompt neurinos due to RQPM or QGSM  and the conventional ones, -- does not contradict to these limits.   

\section{Electron neutrino fluxes and neutrino flavor ratio \label{sec:nue}} 

\begin{figure}[!t] 
 \centering 
\includegraphics[width=78 mm]{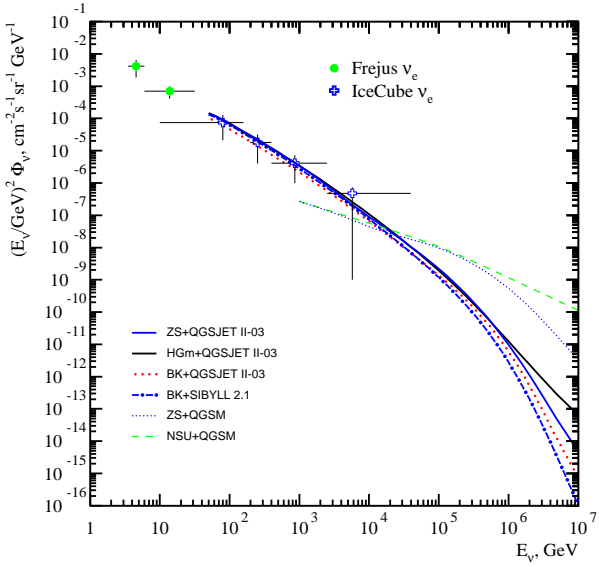} 
\caption{Calculated conventional and prompt  ($\nu_e+\bar\nu_e$)  spectra compared to Frejus~\cite{frejus}  and IceCube data~\cite{ice79_nue}.}
\label{IC_nue}  
\end{figure}
\begin{figure}[!t]
 \centering 
\includegraphics[width=78 mm]{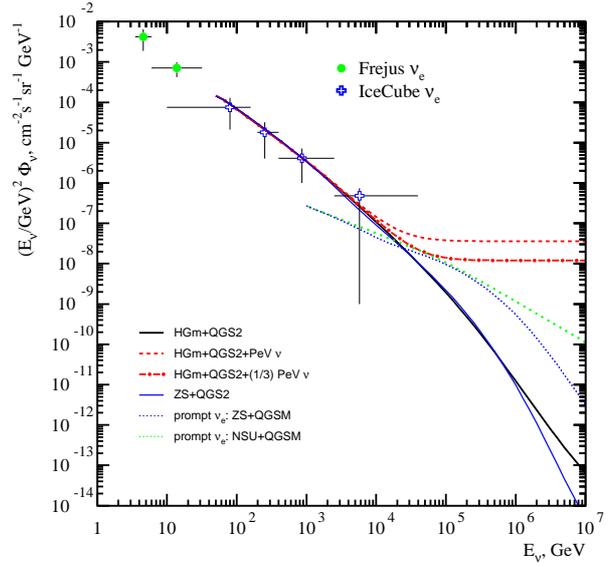} 
\caption{Atmosperic ($\nu_e+\bar\nu_e$) spectrum and diffuse flux of cosmic neutrinos.}
\label{IC_PeVnu}  
\end{figure}

Recently the IceCube  publish results for the electron neutrino spectrum  measured in the energy range $\sim 80$ GeV - $6$ TeV~\cite{ice79_nue}, making possible evaluation the neutrino flavour ratio  and  comparison it with  predictions.  Results of calculation of the atmospheric  ($\nu_e+\bar\nu_e$) flux with  QGSJET II-03 and SIBYLL 2.1 for three parameterisations of cosmic ray spectra are presented in Figs.~\ref{IC_nue} along the measurement data.   
In Fig.~\ref{IC_PeVnu} is  shown also the contribution of diffuse flux (red dashed and dash-dotted lines) of cosmic neutrinos added to the  atmospheric  conventional neutrino flux which are calculated with usage of QGSJET-II-03 for the Gisser spectrum (HiG). Upper dashed red line in this figure depicts the the sum of the atmospheric electron neutrino flux and  the diffuse flux of cosmic neutrinos with an the $E^{-2}$-spectrum, 
$E_\nu^2\phi_\nu=3.6\cdot 10^{-8}$, (cm$^2$\,s\,sr)$^{-1}$ GeV (assuming a flavor ratio of $\nu_e:\nu_\mu:\mu_\tau=1:0:0$), dash-dotted red line corresponds to sum of atmospheric electron neutrinos  and astophysical ones, $E_\nu^2\phi_\nu=1.2\cdot 10^{-8}$, (cm$^2$\,s\,sr)$^{-1}$ GeV, for  the 
 assumption $\nu_e:\nu_\mu:\mu_\tau=1:1:1$). This limit obtained by IceCube~\cite{IC_dif2011} for the energy range above $1$ PeV is compatible with the two PeV neutrino events~\cite{pevnu2013} with energies $1.04 \pm 0.16$ and $1.14 \pm 0.17$ PeV which were detected by the IceCube neutrino telscope.
\begin{figure}[!b]  
  \centering  
\includegraphics[width=75 mm] {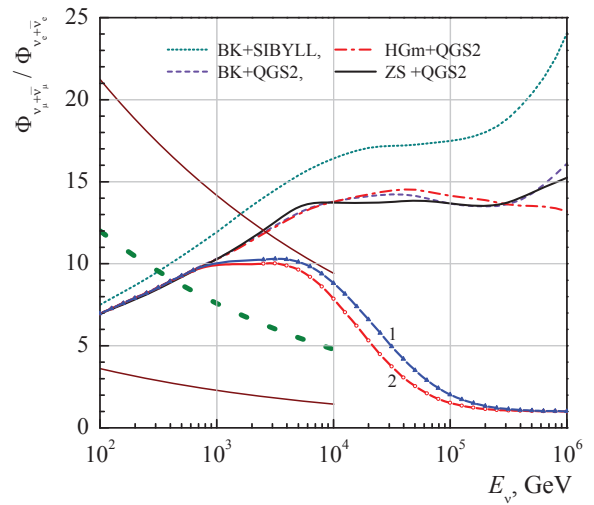} 
 \caption{Neutrino flux ratio ($\nu_\mu+\bar\nu_\mu)/(\nu_e+\bar\nu_e$).} 
  \label{flavor-1} 
 \end{figure}
 
Since IceCube has measured energy spectra both of muon and electron neutrino, we may try to construct the neutrino flavour ratio $(\nu_\mu+\bar\nu_\mu)/(\nu_e +\bar\nu_e)$ and  check for agreement the calculations  with experimental data. 	
The conventional neutrino flavour ratio, $\phi_{\nu_\mu+\bar\nu_\mu}/\phi_{\nu_e+\bar\nu_e}$, calculated for different parameterisations of cosmic ray spectra, as it is seen in Fig.~\ref{flavor-1}, is more sensitive to hadronic models than to the primary spectrum. The difference of neutrino flux predictions related to choice of hadronic models  is clearly seen: curves display the scale of difference between the conventional ($\nu_\mu+\bar\nu_\mu$) and  ($\nu_e+\bar\nu_e$ spectra), calculated with usage of QGSJET II, SIBYLL  for all of the 3  PCR models, HGm, ZS  and BK.  Monte Carlo calculations give the high ratio, $25-30$ (see Fig. 2 in~\cite{fedyn12}), unlike to that of present work, $10-16$.

Thick dashed line (green) corresponds to the IceCube data, the solid curves above and below the green one signify a rough estimate of the IceCube data uncertainties.  
Curves marked as 1 and 2 in Fig.~\ref{flavor-1}  depict the total neutrino flavor ratio comprising the conventional, prompt and astrophysical neutrinos (these last were added in line with  with  two assumptions indicated in Fig~\ref{IC_PeVnu}).
Thus one may assume that IceCube atmospheric neutrino measurement data give an indication
that astrophysical electron neutrinos are probably visible in the energy region $1-10$ TeV.  
There is obscure behavior of the flavor ratio in the range  $100$ GeV - $1$ TeV, probably related to  neutrino oscillations: the oscillation signal was detected within a low-energy muon neutrino sample ($20-100$ GeV) in data collected by DeepCore~\cite{dcIC13}. 

\section{Summary} 
The calculations of the high-energy atmospheric neutrino flux demonstrate rather weak dependence on the primary specrtum models in the energy range $10^2-10^5$ GeV.  However the picture appears less steady because of sizable flux differences originated from  the models of high-energy hadronic interactions. As it can be seen by the example of the models QGSJET-II-03 and SIBYLL 2.1, the major factor of a discrepancy in the conventional neutrino flux is the kaon production in nucleon-nucleus collisions.  

Calculated spectra of muon neutrinos  show  apparent dependence on the spectrum model and composition of primary cosmic rays in the range above  $100$ TeV, which includes the ``knee''. 
Also in this region uncertainties appear due to production cross sections of charmed particles.
The prompt neutrino contribution due to QGSM leads to better agreement with the IceCube data above $100$ TeV. The total flux of the conventional and prompt neutrinos calculated with QGSJET II-03 and QGSM describes the IceCube  data well enough. The QGSM predicted muon neutrino flux  in the range $200-400$ TeV as well as the RQPM  one does not violate the upper limit on the diffuse flux of astrophysical neutrinos obtained by IceCube59~\cite{ice59_lim}.

However, observation of the PeV-energy neutrino events by IceCube~\cite{pevnu2013} 
changes  drastically  the situation concerning the prompt neutrino contribution.
If the first indication of astrophysical neutrinos by IceCube would be confirmed in further studies (particularly in the measurement of electron neutrino flux above $10$  TeV),  then 
the  atmospheric $\nu_e$  flux  uncertainty due to the charm production becomes negligible at the energy above $10$ TeV (see Fig.~\ref{IC_PeVnu}).  

Neutrino flavor ratio extracted from IceCube data  at energies up to $5.7$ TeV  decreases in the energy range $0.1 - 5$ TeV  energy  contrary to the conventional neutrino flux computation (Fig.~\ref{flavor-1}).
Presumbly there are three reasons of such behavior: i) the neutrino oscillations in low-energy  part of this  energy range, ii) early arising contribution from decays of charmed particle (the prompt neutrinos),  and at last iii) increasing with energy contribution of the diffuse neutrino flux at higher energies.

Preliminary and superficial analysis leads to the assumption: IceCube atmospheric neutrino data  give indication  that astrophysical electron neutrinos should be observable at the energy some $10$ TeV, if the power law $E^{-2}$  is valid for astrophysical neutrino spectrum in this range. 
Whether this optimism has any grounding in reality remains to be seen.  


{
{\footnotesize{\bf Acknowledgments:}  
We thank Prof. V.A. Naumov for useful discussion, Dr. A.A. Kochanov, Ms. O.N. Petrova for the help in the early stage of the work.

Work is supported by the Russian Federation Ministry of Education and Science, 
agreement No.14.B37.21.0785.
} 


\end{document}